\documentstyle[aps,epsbox]{revtex}

\begin{document}
\draft
\title{R-mode oscillations of rapidly rotating Newtonian stars --- 
A new numerical scheme and its application to the spin evolution
of neutron stars}
%
%
\author{Shigeyuki Karino$^1$, Shin'ichirou Yoshida$^1$, 
Shijun Yoshida$^2$, and Yoshiharu Eriguchi$^1$}
\address{$ ^1$ Department of Earth Science and Astronomy,\\
         Graduate School of Arts and Sciences,\\
         University of Tokyo, Komaba, Meguro, Tokyo 153-8902, Japan}
%
%
\address{$ ^2$ Astronomical Institute, Graduate School of Science,\\
         Tohoku University, Sendai 980-8578, Japan}

\date{\today}
\maketitle
\begin{abstract}
%
%

We have developed a new numerical scheme to solve r-mode oscillations of 
{\it rapidly rotating polytropic stars} in Newtonian gravity.  In this scheme, 
Euler perturbations of the density, three components of the velocity are 
treated as four unknown quantities together with the oscillation frequency. 
For the basic equations of oscillations, the compatibility equations are used 
instead of the linearized equations of motion.

By using this scheme, we have solved the classical r-mode oscillations 
of rotational equilibrium sequences of polytropes with the polytropic indices
$N = 0.5, 1.0$ and $1.5$ for $m = 2, 3$ and $4$ modes. Here $m$ is the rank
of the spherical harmonics $Y_l^m$.  These results have been applied to 
investigate evolution of uniformly rotating hot young neutron stars by 
considering the effect of gravitational radiation and viscosity.
We have found that the maximum angular velocities of neutron stars
are around 10-20\% of the Keplerian angular velocity irrespective of
the softness of matter. This confirms the results obtained from the
analysis of r-modes with the slow rotation approximation employed by many 
authors.

\end{abstract}

\widetext

%
\section{Introduction}

About 110 years ago, Bryan analyzed oscillations of self-gravitating and 
spheroidal liquids and obtained the period equations for oscillation modes 
whose frequencies are proportional to the angular velocity~\cite{bryan1889}.  
These modes are similar to those which were investigated for rotating 
realistic compressible stars and named r-modes by Papaloizou \& 
Pringle~\cite{pp78} (see also e.g. \cite{pbr81,saio82}).

It is only very recently that the r-mode has been rediscovered from  a 
standpoint of its strong instability and intensively studied.
For sufficiently rapidly rotating stars whose oscillations
are coupled to gravitational radiation, instability induced by 
gravitational radiation was discovered by Chandrasekhar~\cite{c70} and by 
Friedman \& Schutz~\cite{fs78}.
Thus it is called the CFS instability 
(for reviews, see e.g. \cite{fi92,l97a,l97b,fl99}).  
Andersson~\cite{a98} has pointed out that the r-mode oscillations 
suffer from the CFS instability and play a very important role in a 
first few years after formation of neutron stars.  

In subsequent studies by many authors 
(e.g. \cite{fm98,lom98,li99,lf99,aks99,yl99}), it has become clear that the 
r-mode oscillations make newly born neutron stars significantly unstable by 
gravitational radiation emission for a temperature range around $10^{9}$ K.  
For the CFS instability of the f-mode oscillations, the effect of viscosity 
stabilizes neutron stars almost all range of the temperature.
Consequently the angular velocity would be affected only slightly
and would become 90-95\% of the Keplerian angular velocity.

Contrast to the CFS instability of f-modes, the CFS instability of r-modes
will not be stabilized for a certain range of the temperature $T$ of neutron 
stars, i.e., roughly around $T \sim 10^{8 \sim 10}$~K, even if there 
exists viscosity.  
This may be explained by the weakness of bulk viscosity damping
because the volume expansion rate of the r-mode oscillation is small.
Also the critical rotational frequency of the star from which the
instability sets in is zero, in contrast to the case of the f-modes.
It leads to the enhancement of the instability as well.
Due to the effect of the r-mode instability, neutron stars which have been 
born with $T \sim 10^{11}$~K will lose most of their angular momentum in only 
one year and settle down to slowly rotating configurations. This time scale 
is determined from that of cooling of newly born neutron stars down to $T 
\sim 10^{9}$~K. 
Thus there exist critical angular velocities for neutron 
stars above which no stable configurations exist
for each value of the temperature.

This sets a severe limit to the scenario of evolution of newly born neutron 
stars, in particular, to that of millisecond pulsars. From the observational 
point of view, the rotational periods of neutron stars were considered to be 
rather long (see e.g. \cite{na87,lbdh93}), although from the angular momentum
considerations of the progenitors of neutron stars they could be shorter 
than $\sim 10$ ms.  Therefore, some mechanism which sets limit to the 
rotational periods of neutron stars has been required and looked for. From 
the significant feature of the r-mode instability, it could be the promising 
one which explains the observational results. 

However, we should note that almost all results mentioned above have been 
obtained from stability analysis of slowly rotating stars except two works by 
Bryan~\cite{bryan1889} and Lindblom \& Ipser~\cite{li99}.  
For example, Lockitch \& Friedman~\cite{lf99} have introduced the slow 
rotation approximation to get eigenfrequencies and eigenfunctions of the 
classical r-modes and generalized r-modes for polytropic stars (about 
generalized r-modes, see also \cite{li99}), and Andersson, Kokkotas, 
\& Schutz~\cite{aks99} 
have also analyzed slowly rotating equilibrium configurations to estimate the 
spin evolution of young neutron stars. As for the results of 
\cite{bryan1889} and those of \cite{li99}, Maclaurin 
spheroids were analyzed.  Since realistic stars are not 
incompressible fluid and newly born 
neutron stars might be rotating with very short periods, it would be 
desirable to investigate r-mode oscillations of {\it rapidly rotating} and 
{\it compressible} stars. 

In the previous Letter paper~\cite{ykye99}, we have shown some representative 
results of the classical r-mode oscillations of rapidly rotating polytropic 
stars with the polytropic index $N = 1$.  Those results have been obtained by 
developing a new scheme to handle r-mode oscillations of rapidly rotating 
polytropic stars in Newtonian gravity.  In this paper, we will explain the 
new scheme in detail and show classical r-mode oscillations of rotational
equilibrium configurations for a wide range of $N$ and $m$.  

The numerical scheme is an extended version of the one which was used to 
analyze f-mode instability by Yoshida \& Eriguchi~\cite{ye95}. 
After computing r-mode oscillations for equilibrium sequences of 
several rotating polytropes, we will calculate time scales of evolution
due to dissipation by taking account of the effect of viscosity
and gravitational radiation and obtain critical curves for the r-mode
instability in the temperature -- rotational frequency plane by following the 
analysis method proposed by Ipser \& Lindblom~\cite{il91}.  
In the same plane, we can also draw ``evolutionary curves" 
due to cooling by using the standard modified 
URCA process\cite{st83} or other mechanisms.  

Our newly developed scheme for the oscillations of rotating stars will be 
explained in Section 2. In Section 3, we will show our computational results 
for the classical r-mode oscillations of rotating polytropes. Time scales
of evolution due to gravitational radiation as well as viscosity will be 
calculated by using the eigenfrequencies and eigenfunctions of r-mode
oscillations. By using these time scales we will be able to get
critical curves for the r-mode instability and the `evolutionary' curves 
for hot young neutron stars.
Finally, in Section 4, we will discuss the uncertain issues included
in our analysis, and some other important problems which we 
will have to treat carefully in the future.

\section{Scheme of linear perturbations for uniformly rotating polytropes}

\subsection{Equilibrium states}

In this paper we will treat {\it uniformly rotating polytropes} in the 
framework of {\it Newtonian} gravity.  Axisymmetric equilibrium configurations
of rapidly rotating Newtonian stars are obtained by solving the hydrostatic 
equation, Poisson's equation and the equation of state as follows:
\begin{equation}
\frac{1}{\rho_0}\nabla p_0
=-\nabla \phi_0 + r \sin \theta \Omega ^2 \vec{e}_R \ ,
\end{equation}
\begin{equation}
\triangle \phi_0
=4 \pi G \rho_0  \ ,
\end{equation}
\begin{equation}
p_0 = K \rho_0^{1+\frac{1}{N}} \ ,
\end{equation}
where $p_0$, $\rho_0$, $\phi_0$, $\Omega$, $\vec{e}_R$, $G$, $K$ and $N$
are the density, the pressure, the gravitational potential, the angular 
velocity, a unit vector in 
the R direction of the cylindrical polar coordinate ($R, \varphi, z$), 
the gravitational constant,
the polytropic constant, and the polytropic index, respectively. 
Through this paper, the spherical polar coordinates $(r, \theta, \varphi)$ 
are used.

In order to handle properly the boundary condition of the potential at 
infinity, the potential is transformed into the integral representation
by using the Green's function. These equations are solved by applying the 
method developed in \cite{em85}.

In actual computations we have obtained three polytropic equilibrium 
sequences with $ N = 0.5, 1.0$, and $1.5$.  The mesh number in
this paper is $(r \times \theta) = (32 \times 11)$.  Here equidistant mesh
sizes in $r$- and $\theta$-directions are used.

\subsection{Linear perturbations of uniformly rotating and 
axisymmetric polytropes}

Once equilibrium configurations are obtained, we can proceed to 
analysis of linear perturbations of equilibrium states. 
Basic equations for the perturbed states can be written as follows.
We will express Eulerian perturbations by using $\delta$, say $\delta g$
for the physical quantity $g$.  Since in this paper rotating polytropes 
will be employed, the linearized equations need to be treated separately 
for $N = 0$ and $N \ne 0$ cases. 

\subsubsection{Basic equations for $N \ne 0$ polytropes}

For $N \ne 0$ polytropes, the linearized equations of the continuity 
and the linearized equations of motion are written as follows:
\begin{eqnarray}
{1 \over \rho_0} {\partial \delta \rho_1 \over \partial t} +
\frac{\partial \delta u_1}{\partial r}
+\frac{2}{r}\delta u_1
+\frac{1}{\rho_0}\frac{\partial \rho_0}{\partial r}\delta u_1
+\frac{1}{r}\frac{\partial \delta v_1}{\partial \theta}
+\frac{1}{r}\frac{1}{\rho_0}\frac{\partial \rho_0}{\partial \theta}
\delta v_1  \nonumber \\
+\frac{1}{r}\cot \theta \delta v_1
+ \Omega\frac{1}{\rho_0} {\partial \delta \rho_1 \over \partial \varphi}
+ \frac{1}{r\sin\theta} {\partial \delta w_1 \over \partial \varphi}
= 0 \ , \label{eq:contorg1}
\end{eqnarray}

\begin{equation}
{\partial \delta u_1 \over \partial t}
+\left(1+\frac{1}{N}\right)\frac{\partial}{\partial r}(\rho_0^{\frac{1}{N}-1}
\delta \rho_1)
+ \frac{\partial \delta \phi_1}{\partial r}
+ \Omega {\partial \delta u_1 \over \partial \varphi}
- 2 \Omega \sin \theta \delta w_1
=  0 \ ,
\end{equation}
\begin{equation}
{\partial \delta v_1 \over \partial t}
+\left(1+\frac{1}{N}\right)\frac{1}{r}\frac{\partial}{\partial \theta}(\rho_0^{\frac{1}
{N}-1}\delta \rho_1)
+ \frac{1}{r}\frac{\partial \delta \phi_1}{\partial \theta}
+ \Omega {\partial \delta v_1 \over \partial \varphi}
- 2\Omega \cos \theta \delta w_1
= 0 \ ,
\end{equation}
\begin{eqnarray}
{\partial \delta w_1 \over \partial t} +
\left(1+\frac{1}{N}\right)\frac{1}{r \sin \theta} {\partial (\rho_0^{\frac{1}{N}-1}
\delta \rho_1) \over \partial \varphi}
+ \frac{1}{r\sin\theta} {\partial \delta \phi_1 \over \partial \varphi} 
+ 2\Omega\sin \theta \delta u_1  \nonumber \\
+ 2\Omega \cos \theta \delta v_1
+ \Omega {\partial \delta w_1 \over \partial \varphi}
=  0.  \label{eq:hydororg1}
\end{eqnarray}
Here perturbed quantities of the density, three components of the
velocity, $(u, v, w)$ in the spherical coordinates, the gravitational 
potential are expressed as
$\delta \rho_1, \delta u_1, \delta v_1, \delta w_1$ and
$\delta \phi_1$, respectively. We have made use of the relation between the 
perturbed pressure, $\delta p_1$, and the perturbed density as follows:
\begin{equation}
\delta p_1 = \left(1 + {1 \over N}\right) K \rho_0^{1/N} \delta \rho_1 \ . 
\end{equation}

Since the unperturbed states are axisymmetric and stationary,
all perturbed quantities can be expressed by using the expansion
as follows:
\begin{eqnarray}
  \delta \rho_1(r, \theta, \varphi, t) &=& \sum_m \exp(i(\sigma t - m \varphi))
   \rho_m(r,\theta) \ , \nonumber \\
  \delta u_1(r, \theta, \varphi, t) &=& \sum_m \exp(i(\sigma t - m \varphi))
   u_m(r,\theta) \ , \nonumber \\
  \delta v_1(r, \theta, \varphi, t) &=& \sum_m \exp(i(\sigma t - m \varphi))
   v_m(r,\theta) \ ,  \\
  \delta w_1(r, \theta, \varphi, t) &=& \sum_m \exp(i(\sigma t - m \varphi))
   w_m(r,\theta) \ , \nonumber \\
  \delta \phi_1(r, \theta, \varphi, t) &=& \sum_m \exp(i(\sigma t - m \varphi))
   \phi_m(r,\theta) \ , \nonumber
\end{eqnarray}
where $\sigma, \rho_m, u_m, v_m, w_m$ and $\phi_m$ are the oscillation 
frequency and the expansion coefficients of corresponding quantities, 
respectively.

When we define the physical quantities
\begin{eqnarray}
\delta \rho  & \equiv & \rho_m \ , \nonumber \\
\delta u_r  & \equiv & i u_m \ , \nonumber \\
\delta v_{\theta}  & \equiv & i  v_m \ ,  
\label{quantity} \\
\delta w_{\varphi}  & \equiv & w_m \ , \nonumber \\
\delta \phi  & \equiv & \phi_m \ , \nonumber
\end{eqnarray}
and apply the above expansion, the perturbed equations become as follows:
\begin{eqnarray}
\frac{\partial \delta u_r}{\partial r}
+\frac{2}{r}\delta u_r
+\frac{1}{\rho_0}\frac{\partial \rho_0}{\partial r}\delta u_r
+\frac{1}{r}\frac{\partial \delta v_{\theta}}{\partial \theta}
+\frac{1}{r}\frac{1}{\rho_0}\frac{\partial \rho_0}{\partial \theta}\delta v_{\theta}  \nonumber \\
+\frac{1}{r}\cot \theta \delta v_{\theta}+m\Omega\frac{1}{\rho_0}\delta \rho+\frac{m}{r\sin\theta}\delta w_{\varphi}
= \sigma \frac{1}{\rho_0}\delta \rho, \label{eq:cont1}
\end{eqnarray}
\begin{equation}
-\left(1+\frac{1}{N}\right)\frac{\partial}{\partial r}(\rho_0^{\frac{1}{N}-1}\delta \rho)
- \frac{\partial \delta \phi}{\partial r}
- m\Omega\delta u_r
+2\Omega \sin \theta \delta w_{\varphi}
=\sigma \delta u_r,
\label{fr-n}
\end{equation}
\begin{equation}
-\left(1+\frac{1}{N}\right)\frac{1}{r}\frac{\partial}{\partial \theta}(\rho_0^{\frac{1}
{N}-1}\delta \rho)
-\frac{1}{r}\frac{\partial \delta \phi}{\partial \theta}
+m\Omega\delta v_{\theta}
+2\Omega \cos \theta \delta w_{\varphi}
=\sigma \delta v_{\theta},
\label{fth-n}
\end{equation}
\begin{equation}
\left(1+\frac{1}{N}\right)\frac{m\rho_0^{\frac{1}{N}-1}\delta\rho}{r \sin \theta}
+\frac{m}{r\sin\theta}\delta\phi
+2\Omega\sin \theta \delta u_r
+2\Omega \cos \theta \delta v_{\theta}
+m\Omega\delta w_{\varphi}
=\sigma\delta w_{\varphi}.\label{eq:hyd1}
\end{equation}

\subsubsection{Basic equations for $N = 0$ polytropes}

For $N=0$ polytropes, we can follow the same procedure as
that for $N \ne 0$ polytropes by choosing the pressure
perturbation as one of the unknown quantities instead of the
density perturbation. The final equations can be written as
follows:
\begin{equation}
\frac{\partial \delta u_r}{\partial r}
+\frac{2}{r}\delta u_r
+\frac{1}{r}\frac{\partial \delta v_{\theta}}{\partial \theta}
+\frac{1}{r}\cot \theta \delta v_{\theta}
+\frac{m}{r\sin\theta}\delta w_{\varphi}
=0, \label{eq:cont2}
\end{equation}
\begin{equation}
-\frac{1}{\rho_0}\frac{\partial \delta p}{\partial r}
-\frac{\partial\delta\phi}{\partial r}
+m\Omega\delta u_r
+2\Omega \sin\theta \delta w_{\varphi}
=\sigma\delta u_r,
\label{fr-0}
\end{equation}
\begin{equation}
-\frac{1}{\rho_0}\frac{1}{r}\frac{\partial \delta p}{\partial \theta}
-\frac{1}{r}\frac{\partial\delta\phi}{\partial \theta}
+m\Omega\delta v_{\theta}
+2\Omega \cos\theta \delta w_{\varphi}
=\sigma\delta v_{\theta},
\label{fth-0}
\end{equation}
\begin{equation}
\frac{1}{\rho_0}\frac{m\delta p}{r\sin\theta}
+\frac{m}{r\sin\theta}\delta\phi
+2\Omega \sin \theta \delta u_r
+2\Omega \cos \theta \delta v_{\theta}
+m\Omega\delta w_{\varphi}
=\sigma\delta w_{\varphi}.\label{eq:hyd2}
\end{equation}

\subsubsection{Compatibility conditions}

Instead of solving the basic equations derived above, we make use of
the compatibility relations among the equations of motion, in particular,
the compatibility equations between Eqs.~(\ref{fr-n}) and (\ref{eq:hyd1})
and between Eqs.~(\ref{fr-n}) and (\ref{fth-n}) for $N \ne 0$ models or 
those between Eqs.~(\ref{fr-0}) and (\ref{eq:hyd2}) and 
between Eqs.~(\ref{fr-0}) and (\ref{fth-0}) for $N = 0$ models.
They can be written both for $N \ne 0$ and $N = 0$ cases as follows:
\begin{eqnarray}
\frac{2 r \Omega \sin^2 \theta}{m}\frac{\partial\delta u_r}{\partial r}
&+&\frac{2 r \Omega \sin\theta \cos\theta}{m}
\frac{\partial\delta v_{\theta}}{\partial r}
+\frac{r \sin\theta}{m}(m\Omega-\sigma)
\frac{\partial\delta w_{\varphi}}{\partial r} \nonumber\\ 
&&+\left[(m\Omega-\sigma)+\frac{2}{m}\Omega \sin^2\theta\right]\delta u_r
+\frac{2\Omega}{m}\sin\theta \cos\theta\delta v_{\theta}
+(3\Omega-\frac{\sigma}{m})\sin\theta \delta w_{\varphi}
=0,
\label{fr-fphi}
\end{eqnarray}
\begin{eqnarray}
r(m\Omega-\sigma)\frac{\partial\delta v_{\theta}}{\partial r}
+2r\Omega \cos\theta\frac{\partial\delta w_{\varphi}}{\partial r}
-(m\Omega-\sigma)\frac{\partial \delta u_r}{\partial \theta}
-2\Omega \sin\theta\frac{\partial\delta w_{\varphi}}{\partial \theta}
\nonumber \\
+(m\Omega-\sigma)\delta v_{\theta}
=0.
\label{fr-fth}
\end{eqnarray}

\subsubsection{Perturbed gravitational potential}

The perturbed gravitational potential can be expressed by using
the integral form as follows:
\begin{eqnarray}
\delta \phi=-4\pi G \sum_{n,m} \int_0^{\frac{\pi}{2}}
d\theta ' \sin \theta ' \frac{(n-m)!}{(n+m)!} 
P^m_n (\cos \theta) P^m_n (\cos \theta ') \nonumber \\
\times \int_0^{r_s(\theta ')}d r 'r ^{\prime 2} f_n (r,r ')
\delta\rho \nonumber \\
-4\pi G \sum_{n,m} \int_0^{\frac{\pi}{2}}
d\theta ' \sin \theta ' \frac{(n-m)!}{(n+m)!} 
P^m_n (\cos \theta) P^m_n (\cos \theta ') \nonumber \\
\times f_n (r,r_s(\theta ')) \rho_0(r_s(\theta '),\theta ')
\delta r_s (\theta ') \ ,
\label{potential}
\end{eqnarray}
where $r_s (\theta)$ is the surface radius of the equilibrium configuration
and $\delta r_s$ is the change of the surface. Here, functions $f_n(r,r')$
are defined as 
%

\begin{equation}
f_n (r,r') = \left\{ \begin{array}{ll}
\frac{1}{r}( \frac{r'}{r})^n & \quad (r'< r),\\
\frac{1}{r'}( \frac{r}{r'})^n & \quad (r'\geq r),
\end{array}
\right.
\end{equation}
and $P_n^m$'s are associated Legendre polynomials. 

It is noted that the boundary conditions for the gravitational 
potential, i.e. regularity condition throughout the space and 
the flatness at infinity, are automatically included in
this integral representation.

\subsubsection{Boundary conditions on the stellar surface}

Since the perturbed gas must flow along the perturbed surface,
the boundary condition for the flow velocity on the surface 
(for $N \ne 0$ configurations) or that for the surface (for $N = 0$ 
configurations) can be expressed as follows:
\begin{equation}
\frac{\partial\rho_0}{\partial r}\delta u_r
+\frac{1}{r} \frac{\partial\rho_0}{\partial \theta}
\delta v_\theta+(m\Omega-\sigma)\delta\rho=0 \ ,
\label{boundary-n}
\end{equation}
for $N \not = 0$ polytropes, or 
\begin{equation}
-\delta u_r+\frac{\delta v_{\theta}}{r_s}
\frac{dr_s}{d\theta}+(m\Omega-\sigma)\delta r_s=0 \ ,
\label{boundary1-0}
\end{equation}
for $N=0$ polytropes. 

For $N=0$ polytropes, since the unknown quantity $\delta r_s$ has
to be solved in addition to other physical quantities, 
we need to include the following relation which is
satisfied on the surface:
\begin{equation}
\delta p_s=-\left.\left(\frac{\partial p_0}{\partial r}\right)\right|_s \delta r_s
=\rho_0\left.\left[\left(\frac{\partial \phi_0}{\partial r}\right)-r_s \sin^2\theta\Omega^2\right]\right|_s
\delta r_s \ .
\label{boundary2-0}
\end{equation}

\subsubsection{Solving scheme}

Our basic equations and boundary conditions 
are Eqs.~(\ref{eq:cont1}), (\ref{eq:hyd1}), (\ref{fr-fphi}), (\ref{fr-fth}),
and (\ref{boundary-n}) for $N \ne 0$ polytropes and
Eqs.~(\ref{eq:cont2}), (\ref{eq:hyd2}), (\ref{fr-fphi}), (\ref{fr-fth}), 
(\ref{boundary1-0}), and (\ref{boundary2-0}) for $N = 0$ polytropes,
although we will not show the results for $N = 0$ polytropes
in this paper.

In order to handle the surface boundary conditions as precisely as possible, 
we introduce surface--fitted coordinates as follows (see e.g.
\cite{em85,em91,ue94}):
\begin{eqnarray}
r^*      & \equiv & {r \over r_s(\theta)} \ , \\
\theta^* & \equiv & \theta \ .
\end{eqnarray}

We choose equidistantly spaced mesh points on the $r^*$-  and 
$\theta^*$-coordinates. The basic equations expressed in the new coordinate 
system are discretized on these mesh points. The derivatives with respect 
to $r^*$ and $\theta^*$ are replaced by the central difference scheme with 
the second order accuracy for the most mesh points except for those near 
the surface and the equatorial plane.  The derivatives with respect to $r^*$ 
near the surface and those with respect to $\theta^*$ near the equator are 
replaced by difference schemes with higher order accuracy.

Instead of solving the eigenvalue problem directly, we add one more 
condition to the basic equations of the system and solve for all the unknown
quantities simultaneously by using the Newton-Raphson iteration scheme.
Since we focus only on the classical r-mode in this paper, 
we adopt the following condition as the additional equation:
\begin{equation}
\delta v_{\theta} = 1.0  \ , \ \ \ \hbox{\rm on the equatorial surface} \ .
\end{equation}

\section{Numerical Results}

We have computed equilibrium sequences of uniformly rotating polytropes with
$N = 0.5, 1.0$ and $1.5$ by starting from spherical models to terminal models 
which are configurations just before shedding mass from the equatorial 
surface. By using these equilibrium models, classical r-mode oscillations 
for $m = 2, 3$ and $4$ modes have been solved and eigenfrequencies and 
eigenfunctions have been obtained.  In the actual numerical computations, 
the mesh number in $(r^*, \theta^*)$ coordinates is the same as that used 
in obtaining the equilibrium configurations, i.e. $(r^* \times \theta^*) 
= (32 \times 11)$.

\subsection{Eigenfrequencies and eigenfunctions of the r-mode oscillations}

In Figs.~1--3, the eigenvalue of the $m=2$ r-mode is plotted against the 
dimensionless angular velocity of the equilibrium configurations 
for $N = 1.0, 1.5.$ and $0.5$ polytropes, respectively. In these figures,
the eigenfrequency $\sigma$ is normalized by using the angular velocity 
$\Omega$ and the angular velocity is normalized as 
$\Omega /\sqrt{4\pi G \rho_c}$.  Here $\rho_c$ is the central density of the 
star. 

Since classical r-mode oscillations of polytropes have been investigated
by using the slow rotation approximation to the third order of the angular 
velocity in~\cite{yl99}, we can compare our results with theirs.
As seen from these figures except Fig.~3, the relative errors between the 
numerical results and those of the approximate results are 0.4\% for the 
slowest configurations. It is natural that for rather rapid rotational 
models difference becomes larger because the assumption of slow rotation 
is violated.

Fig.~4 shows the eigenfrequencies of $N=1.0$ rotating polytropes for different 
values of $m$. From this figure, it is clear that the behaviors of the 
eigenfrequency along the rotational sequence for different values of $m$ 
are almost the same.  This is also found for $N = 0$ incompressible 
configurations, i.e. for Maclaurin spheroids \cite{li99}.
In Fig.~5, the eigenfrequencies of $N = 0.5, 1.0$ and $1.5$ polytropes
for $m = 3$ perturbations are plotted.

In Figs.~6--7, we have plotted the profile of the perturbed quantities,
($\delta u_r$, $\delta v_{\theta}$, $\delta w_{\varphi}$, $\delta \rho$), 
against the radial distance normalized by the stellar radius, i.e. $r^*$. 
Figs.~6 and 7 display the distributions of the r-mode oscillation of the 
polytropes with $N = 1.0$ for $r_p/r_e = 0.99 $ and $r_p/r_e = 0.71 $
models, respectively.  Here $r_p$ and $r_e$ are the polar radius and the
equatorial radius, respectively.

\subsection{CFS instability}

In order to analyze the stability due to dissipative mechanisms such as
gravitational radiation and viscosity, there are several schemes to
include the dissipative forces. A direct method is to solve the
perturbed equations of motion in which the forces due to gravitational
radiation and viscosity are taken into account \cite{ye95}. Another
method is to evaluate the time scales of the system change due to 
gravitational radiation and viscosity \cite{il91} 
(see also \cite{l97a,l97b,fl99}).  Although two methods are essentially 
the same, the formulation of the latter method is very simple.
Thus in this paper we will follow the Ipser-Lindblom method.

In their method, the time derivative of the energy of the oscillation
mode plays an important role.  The energy of the mode, $E(t)$, as
measured in the rotating frame can be expressed as
\begin{equation}
E(t) = {1 \over 2} \int \left[ \rho_0 \delta v^a \delta v_a^*
 + \left( {\delta p \over \rho_0} - \delta \phi \right) \delta \rho^*
\right] d^3 x \ ,
\end{equation}
where $^*$ represents complex conjugation.  From this definition and
by using the perturbed continuity, the perturbed equations of motion with
the dissipative forces and the perturbed Poisson's equation, we can get
the following expression for the time derivative of the mode energy
in the rotating frame:
\begin{equation}
\frac{dE}{dt} \equiv \frac{dE_{gr}}{dt} + \frac{dE_s}{dt} + \frac{dE_b}{dt}
\ ,
\label{eq:dEgdt}
\end{equation}
where
\begin{eqnarray}
\frac{dE_{gr}}{dt} & \equiv & 
- \sigma (\sigma-m\Omega)\sum_{l \ge 2, l\ge m} N_l 
\sigma^{2l} \left(\left|\delta D_{lm}\right|^2 
+ \left|\delta J_{lm}\right|^2\right) \ , 
\label{energyloss} \\
\frac{dE_s}{dt} & \equiv & - \int 2 \eta \delta \sigma^{ab} 
\delta\sigma_{ab}^* d^3x \ ,  \\
\frac{dE_b}{dt} & \equiv & - \int \zeta \left|\delta\Theta\right|^2 d^3x \ .
\end{eqnarray}
Here $N_l, D_{lm}, J_{lm}, \eta, \zeta, \sigma_{ab}$ and $\Theta$
are a coupling constant, the mass multipole, the mass current multipole, 
the coefficient of shear viscosity, the coefficient of bulk viscosity, 
the shear and the expansion, respectively.  They are defined as:
\begin{eqnarray}
N_l & \equiv & {4 \pi G \over c^{2l+1}} 
    {(l+1)(l+2) \over l(l-1)[(2l+1)!!]^2} \ , \\
D_{lm} & \equiv & \int r^l \rho Y_l^{*m} d^3x \ , \\ 
J_{lm} & \equiv & {2 \over c} {1 \over l+1} \int r^l (\rho \vec{v}) \cdot 
(\vec{r} \times \nabla Y_l^{*m}) d^3x \ , \\ 
\sigma_{ab} & \equiv & \nabla_{(a}v_{b)} - {\Theta \over 3} \delta_{ab} \ , \\
\Theta & \equiv & \nabla_c v^c \ ,
\end{eqnarray}
where $c$ is the speed of light and a round bracket in the subscript means 
symmetrization among indices.

For slowly rotating Newtonian stars, the eigenfrequency to the first order of
the angular velocity is expressed as \cite{pp78,pbr81}
\begin{equation}
\sigma_{\rm slow} = \left(m - \frac{2 m}{l(l+1)} \right) \Omega  \ .
\end{equation}
From this eigenfrequency for slowly rotating Newtonian stars, the 
coefficient of the energy dissipation rate due to gravitational 
radiation Eq.~(\ref{energyloss}), i.e. the first term in the righthand side 
of Eq.~(\ref{eq:dEgdt}), becomes positive as follows:
\begin{equation}
-\sigma_{\rm slow}(\sigma_{\rm slow}-m \Omega) 
= \frac{2m^2(l-1)(l+2)}{l^2(l+1)^2} \Omega^2 > 0 \ .
\end{equation}
Consequently, since the time derivative of the energy dissipation due to
gravitational radiation becomes positive, gravitational radiation makes 
the system unstable by growing the perturbation.

On the other hand, as seen from the second and third terms in the righthand 
side of Eq.~(\ref{eq:dEgdt}), viscosity works as a stabilizing factor for the 
system. Therefore, if we compare the time scale of gravitational radiation and 
that of viscosity, we will be able to conclude the stability of the system.

The time scale of the system change due to dissipation can be
defined as the inverse of the imaginary part of the mode eigenfrequency,
$\tau_r$ as follows: 
\begin{equation}
\frac{1}{\tau_r}
=\frac{\dot{E}_{gr}+\dot{E}_s+\dot{E}_b}{2E} 
\equiv \frac{1}{\tau_{gr}}+\frac{1}{\tau_s}+\frac{1}{\tau_b} \ ,
\label{eq:tau}
\end{equation}
where $\dot{ } = d/dt$.  Positive values of $\tau_r^{-1}$ imply 
that the effect of gravitational radiation emission overcomes the stabilization
effect due to viscosity. In other words, the system is in an unstable state. 
Therefore, the critical condition which divides stable and unstable 
configurations can be defined by the following equation:
\begin{equation}
\frac{1}{\tau_r} = 0 \ .
\label{critical}
\end{equation}

In order to evaluate the time scales we need to know the viscosity coefficients
at high density regions. As for the origin of the viscosity, we adopt the 
same mechanisms as \cite{il91}. The shear viscosity is owing to 
neutron-neutron or electron-electron collisions in the neutron matter 
when the temperature of the star is not so high. On the other hand, for 
high temperature regions, the bulk viscosity which is due to the energy 
generation caused by compression of the neutron star matter because the time 
scale to balance between $\beta$-decay and inverse $\beta$-decay is 
longer than that of the dynamical perturbed motion of the star is dominant. 
The coefficients of each viscosity are written as follows 
\cite{il91,fi79,sa89}:
\begin{eqnarray}
\eta_{nn} & = & 2.0 \times 10^{18} \ \rho_{15}^{9/4} \ T_9^{-2} \ \ 
\hbox{\rm g} \ \hbox{\rm cm}^{-1} \hbox{\rm s}^{-1} \ , \\
\eta_{ee} & = & 6.0 \times 10^{18} \ \rho_{15}^{2} \ T_9^{-2} \ \
\hbox{\rm g} \ \hbox{\rm cm}^{-1} \hbox{\rm s}^{-1} \ , \\
\zeta & = & 6.0 \times 10^{25} \rho_{15}^2 \ (\sigma- m \Omega)^{-2} 
T_9^{6} \ \
\hbox{\rm g} \ \hbox{\rm cm}^{-1} \hbox{\rm s}^{-1} \ , \\
\end{eqnarray}
where 
\begin{eqnarray}
\rho_{15} & \equiv & {\rho \over 10^{15} \ \hbox{\rm g} \ \hbox{\rm cm}^{-3}} 
\ , \\
T_9       & \equiv & {T \over 10^9 \ \hbox{\rm K}} \ .
\end{eqnarray}
For lower temperature regions, we need to consider only electron-electron
collisions for the coefficient of shear.

\subsection{Evolution of hot young neutron stars}

In this paper we will choose the polytropic constant $K$ so that the mass 
and the radius of the star become $M=1.4M_\odot$ and $R \approx 12.5$km in 
the spherical limit. We will call this star a canonical neutron star. 
If we specify the polytropic index, the value of the polytropic constant
$K$ is uniquely determined.

\subsubsection{Critical curve for the r-mode instability}

The critical condition (\ref{critical}) depends on the structure of the
unperturbed configuration, the eigenfunctions of the perturbed state,
the eigenfrequency and the temperature.  Since we assume the polytropic
relation for the canonical neutron stars and consider only classical
r-modes, the structure of the equilibrium state, the 
eigenfunctions and the eigenfrequency are all determined by specifying 
the angular velocity in addition to the polytropic index $N$ and the 
mode number $m$. Thus the critical condition can be considered to be a 
function of the angular velocity and the temperature as well as $N$ and $m$.
In other words, once we specify the values of $N$ and $m$, the critical 
condition can be expressed as a 
{\it critical curve} 
on the 
temperature -- rotational frequency plane.

\subsubsection{`Evolutionary' curve for the stellar spin}

It is not easy to follow realistic evolution of neutron stars by considering
the effect of gravitational loss and viscosity as well as the effect of 
cooling due to neutrino loss. Thus we will choose a different approach.

Once cooling mechanisms are specified, we can estimate cooling time scales 
of the neutron star. In this paper, we assume that the standard modified 
URCA process is dominant in the initial stage of the neutron star
formation \cite{st83}. For this process, the effective cooling time scale 
is a function of the stellar temperature and defined by
\begin{equation}
\tau_{\rm cool}= \left[\frac{d \ln T_9}{dt} \right]^{-1} 
= \frac{6t_c}{T_9^6},
\end{equation}
where $t_c$ is a cooling time for the modified URCA process, 
typically $\sim$ 1y. 

Since the neutron stars evolve due to radiation of gravitational wave and
neutrino loss, we can define a critical state where the time scales
of two mechanisms are the same as follows:
\begin{equation}
\tau_r = \tau_{\rm cool} \ .
\end{equation}
Since this state is also a function of the rotational frequency and the 
temperature if $N$ and $m$ are specified, we can draw an 
{\it `evolutionary' curve of the stellar spin}, whose meaning is made
clear in the following discussion,
on the rotational frequency--temperature plane.

\subsubsection{Spin evolution of canonical neutron stars}

In Figs.~8--12, critical curves for the classical r-modes instability
are drawn in the rotational frequency, $f$, and the temperature plane
for different values of $N$ and $m$.
Above these solid curves, the instability due to gravitational wave emission 
overcomes the stabilizing effect of viscosity and consequently neutron
stars become unstable. 

In these figures, the evolutionary curves of the stellar spin 
are also shown by dashed curves. 
Above these curves, cooling time scale of each cooling mechanism is longer
than the instability time scale. It implies that in this region the neutron
stars evolve nearly downwards by losing the angular momentum but keeping
the temperature constant. In the region below these curves, the stars
sweep leftwards due to neutrino loss by keeping the rotational
frequency almost constant.
 
Suppose we fix a cooling mechanism of the star.
It is certain that neutron stars are born with high temperature with 
$T\sim 10^{11}$ K.  Concerning the rotational frequency of the newly born
neutron stars, they may not rotate so rapidly compared with the rotational
frequency of the Kepler limit \cite{na87,lbdh93} but may rotate with 
certain rotational frequencies above the minimum value of the critical
curve.  It implies that such newly born neutron stars are clearly located 
below the critical curve at the right side of the $T$--$f$ plane.  
Since this region is also below the evolutionary curve,
cooling stars move almost horizontally to leftwards on this plane.
 
During the evolution due to cooling, the stars come to and cross the 
critical curve. Once they come into the regions above the critical
curve, the r-mode instability due to gravitational radiation loss
starts working significantly. As a result, the stars would evolve downwards 
vertically. However, the stars are still located below the evolutionary 
curve and the time scale of cooling is smaller than that of the instability. 
Therefore, due to this cooling effect, the stars move to leftwards almost 
horizontally on the plane. Eventually the stars go across the evolutionary 
curve. As soon as they cross the curve, the instability
works efficiently and the stars move downwards vertically.
If they cross the evolutionary curve once more, the cooling effect 
becomes dominant again and the stars will evolve horizontally. 
Consequently we can consider that the stars evolve almost along the
evolutionary curve of the cooling mechanism being considered. 
In this sense we can regard the evolutionary curves as
curves along which the spin evolution of a cooling star takes place.

As seen from Figs.~8--12, the evolutionary curves are located
above the critical curves. This means that the evolution mentioned above 
will come to an end because they go out of the unstable region of the 
gravitational radiation and reach the lower left part of the $T$--$f$ plane.
Consequently, the neutron stars will settle down to the states with the 
rotational frequencies corresponding to the minimum values of the 
corresponding evolutionary curve for each cooling channel. 

In Figs.~8--10, we show the critical and evolutionary curves for 
$m = 2$, $m=3$ and $m=4$ modes, respectively. From these figures, the 
minimum values of critical curves are about 100Hz, 200Hz, and 300Hz for 
the $m=2$, $m=3$, and $m=4$, respectively.  
Therefore we can see that the instability of the $m = 2$ mode dominates 
other modes with higher values of $m$. 

In Figs.~11--12, we show the critical and evolutionary curves of 
polytropes with $N=0.5$ and $N=1.5$ for the $m=2$ r-mode. These results are 
almost the same as the result of $N=1.0$ polytropes. It means that the 
classical r-mode instability depends little on the compressibility. This 
agrees with the result of \cite{lom98} in which they conclude that the 
instability is insensitive to the equation of state.  In particular, the 
smallest values of the rotational frequency of the critical curves seem 
to be determined uniquely irrespective of the polytropic index as far as the 
value of $m$ is the same. Thus we can conclude that neutron stars
will settle down to states with a unique rotational frequency, i.e.
around 100Hz.

As a summary, neutron stars which are born with high 
temperature ($10^{11}K$) and rather rapid rotational frequency 
(a few times of 100Hz) will be spun down to about 100Hz by the effect 
of the gravitational instability on a quite short time scale.

\section{Discussion}

In this paper we have developed a new numerical method for solving the r-mode 
oscillations of rapidly rotating compressible stars.  By using the obtained
eigenfrequencies and eigenfunctions, we have calculated the condition for
occurrence of the gravitational radiation induced instability. By combining 
the cooling of the neutron stars and the instability, we have confirmed the 
evolutionary scenario of hot young neutron stars proposed from the slow 
rotation approximation, in particular, considerable spin down of the
neutron stars with rather high rotation rates of initial spins on a very 
short time scale (e.g. \cite{lom98,olcsva98}).  The shortness of the spin 
down time scale implies that this mechanism would, effectively, result in 
``birth of slowly rotating neutron stars" because it would be difficult to 
observe the very initial stage with rather high rotational frequencies 
during one or so years just after the formation of neutron stars.  
This might be the reason of observational data of nonexistence of
pulsars with very short initial spin periods \cite{na87,lbdh93}.

It should be noted that our analysis has been carried out under some
restricted assumptions. First, we have formulated the problem in the 
framework of Newtonian gravity. In order to get quantitatively correct
values we have to use general relativity, although it is considerably
difficult to solve oscillations of not only the matter but also 
the gravitational field.  At the moment, no one has succeeded in
treating the perturbations of the gravitational field of axisymmetric
configurations except obtaining neutral points for f-mode oscillations
\cite{sf98}. Concerning the r-mode oscillations in general relativity,
there arises another problem. Since there is the frame dragging,
the effective angular velocity cannot be uniform any more. Some authors suggest
that this may make the problem a singular eigenvalue problem \cite{kj98} (see
also \cite{kh99}).

Second, we have to treat the realistic equation of state for the neutron star 
matter instead of simplified equations of state such as polytropes. However, 
as has been known from the slow rotation approximation \cite{pp78,pbr81,saio82}
and as we have shown for rapidly rotating models in this paper, there arise 
little differences of the oscillations of polytropes with different polytropic 
indices. Therefore, even if we treat the realistic equation of state, the 
situation would not change drastically as far as the classical r-modes
are concerned.

Third, there is some uncertainty about the cooling process.  Although the 
modified URCA process has been taken as the standard mechanism of the neutron 
star cooling, there are some other mechanisms which would contribute to
cooling of neutron stars. Therefore, we have computed evolutionary
curves by introducing several possible cooling mechanisms. Fig.~13 shows 
such curves for several cooling processes: the neutrino 
bremsstrahlung, the quark $\beta$ decay, the pion condensation, and the
direct URCA cooling model, beside the modified URCA process. In our
calculations, the time scale of the direct URCA process is taken 
from \cite{p91} and those of other processes are taken from \cite{st83}
(see also \cite{bp79}). 
As seen from this figure, other cooling mechanisms give obviously different 
evolutionary curves or ``evolutionary paths". Concerning the minimum values
of the evolutionary curves, however, differences among 
different cooling processes are rather small. In other words, the final 
rotational frequencies depend weakly on the cooling mechanisms. 
Consequently the final rotational frequencies would be around
50 $\sim$ 180Hz as far as the scenario mentioned above works for
the hot young neutron stars.

Fourth, the assumption that neutron stars rotate uniformly is also uncertain. 
Since we are dealing with the very beginning of the neutron star formation
stages, newly born neutron stars may rotate differentially, although they
will soon settle down to uniform rotation on a time scale of several years.
The r-mode oscillations of differentially rotating stars are another problem.
There may arise a corotation point within the star and the similar problem
as that in general relativity needs to be solved.

In addition to these problems, we have to discuss the following crucial
issues which may play a more important role in realistic formation and
evolution of neutron stars than the above mentioned things.

The superfluidity is one of the most important but poorly understood physics 
about the realistic neutron stars. Some authors (e.g. \cite{lm95}) have 
suggested that the effect of the superfluid might suppress the r-mode 
instability completely in the low temperature regions, while other authors 
have assumed that this effect might not be strong enough \cite{akst99}
(see also \cite{aks99}). In this paper, 
we assume that the superfluidity will not work effectively. In fact, since the 
realistic neutron star matter cannot be examined by any present experiments 
on the earth, the only way to estimate this effect may be the comparison 
between the theoretical works and the observational data. 

The existence of a solid crust may be another important factor for the r-mode
instability. Bildsten \& Ushomirsky~\cite{bu99} have suggested 
that the existence of the solid crust whose thickness is $\sim$ 1 km 
would work as a stabilizing factor of the r-mode instability and that
the minimum rotational frequency might be 40\% of that of the
Kepler limit. However, it is not clear whether or not the solid crust has 
formed only in a year or so from the formation of the neutron star. 
Furthermore, it is uncertain about the transition region from the
fluid to the solid. 
Thus we have not taken the effect of the crust 
into our considerations in this paper.

The most important factor concerning the neutron star evolution may
be the presence of the magnetic field. There is a high possibility that 
hot young neutron stars might spin down by the effect of the magnetic field
and/or that neutron stars might be born with rather slow rotation rates 
because the angular momentum is transferred by the effect of the
magnetic field from the core to the envelope during the evolution of the
massive stars (e.g. \cite{aks99,na87,lbdh93,sp98}). 
Inclusion of magnetic fields may change the r-mode
characteristics itself.
For instance, magnetic braking may 
enhance the instability
by increasing the amplitude of the mode~\cite{ho-lai}. 
Also suggested is that the 
CFS instability by the Alfv\'en wave radiation may exist
for stars with strong magnetic fields.
On the other hand it is argued~\cite{rezzolla}
that the drifting fluid motion produced by the r-mode
may generate strong toroidal magnetic 
fields. This may suppress the growth of the r-mode amplitude. 
Therefore, the realistic evolution of the neutron stars have to be 
studied by including the r-mode instability as well as the effect of
magnetic fields in the future.



%
%



\begin{figure}
\caption{The eigenfrequency normalized by the angular velocity, 
$\sigma/\Omega$, of the $m = 2$ r-mode is plotted against the dimensionless
angular velocity, $\Omega/\protect\sqrt{4\pi G \rho _c}$, for $N = 1.0$ 
polytropes.
Solid and dashed curves show our numerical result and the approximate result 
obtained in \protect\cite{yl99}, respectively. The relative difference between 
them, at the slowest rotational frequency, i.e. 
$\Omega/\protect\sqrt{4\pi G \rho _c} \approx 0.02$, is 0.4\%.}
\label{eigenfrequency-N1}
\end{figure}



\begin{figure}
\caption{Same as Fig.~1 but for $N=1.5$ polytropes. The sequence terminates
around $\Omega/\protect\sqrt{4 \pi G \rho_c} \sim 0.15$ because it is 
difficult to obtain smooth eigenfunctions for more rapidly rotating 
configurations.}
\label{eigenfrequency-N15}
\end{figure}



\begin{figure}
\caption{Same as Fig.~1 but for $N=0.5$ polytropes.
Since there are no results of the slow rotation approximation 
in \protect\cite{yl99}, only our results are shown.}
\label{eigenfrequency-N05}
\end{figure}



\begin{figure}
\caption{Same as Fig.~1 but for different values of $m$, i.e. 
$m = 2$ (solid curve), $m = 3$ (dashed curve) and $m = 4$ (dotted curve).}
\label{eigenfrequency-m}
\end{figure}

\begin{figure}
\caption{Same as Fig.~1 but for $m = 3$ and for different values of $N$, i.e. 
$N = 0.5$ (dashed curve), $N = 1.0$ (solid curve) and $N = 1.5$ 
(dotted curve).}
\label{eigenfrequency-N}
\end{figure}

\begin{tabular}{cc}
\end{tabular}
\begin{figure}
\caption{The behavior of the eigenfunctions of the $m=2$ r-mode for the $N=1$ 
polytrope with $r_p/r_e = 0.99 $ where $r_p$ and$r_e$ are the polar
radius and the equatorial radius of the surface, respectively. 
Distributions along the $r^*$-direction
of the perturbed quantities are shown: $\delta u_r$ (upper left panel), 
$\delta v_{\theta}$ (upper right panel), $\delta w_{\varphi}$ (lower right 
panel), and $\delta \rho$ (lower left panel).  It is noted that the surface
is located at $r^* = 1.0$.  In each panel, distributions of different 
$\theta$ values are shown by different curves: A curve with No.~L is the
distribution at $\theta = \pi/10*L$.}
\label{eigenfunction-N1-slow}
\end{figure}

\begin{tabular}{cc}
\end{tabular}
\begin{figure}
\caption{Same as Fig.~6 but for the rapidly rotating model
with $r_p/r_e = 0.71 $.
The angular velocity is about 90\% of that of the Kepler limit.}
\label{eigenfunction-N1-rapid}
\end{figure}



\begin{figure}
\caption{Critical rotational frequency for the $m = 2$ r-mode against 
the stellar temperature is plotted by the solid line for $N=1$ polytropes.
This curve is obtained from the condition that $1 / \tau_r = 0$ (see the text
for details).  This solid line divides the space into two distinct 
regions: one for the stable states and the other for the unstable states. 
In addition, the evolutionary curve is also plotted by the dashed line. 
This curve is obtained by equating the dissipation time scale
and the cooling time scale, i.e $\tau_r = \tau_{\rm cool}$.
The cooling time scale is assumed to be due to the modified URCA process.}
\label{critical-N1-m2}
\end{figure}

\begin{figure}
\caption{Same as Fig.~11 but for the $m=3$ mode.}
\label{critical-N1-m3}
\end{figure}

\begin{figure}
\caption{Same as Fig.~11 but for the $m=4$ mode.}
\label{critical-N1-m4}
\end{figure}

\begin{figure}
\caption{Same as Fig.~11 but for the $N=1.5$ polytrope.}
\label{critical-N15-m2}
\end{figure}

\begin{figure}
\caption{Same as Fig.~11 but for the $N=0.5$ polytrope.}
\label{critical-N05-m2}
\end{figure}

\begin{figure}
\caption{Several evolutionary curves are shown for different cooling 
mechanisms. Meanings of the curves are shown in the figure.
The thick solid curve is the critical curve of the $N=1$ 
polytrope for the $m=2$ mode. It should be noted that the smallest values 
of the frequency of evolutionary curves are all located around 100Hz.}
\label{cooling}
\end{figure}

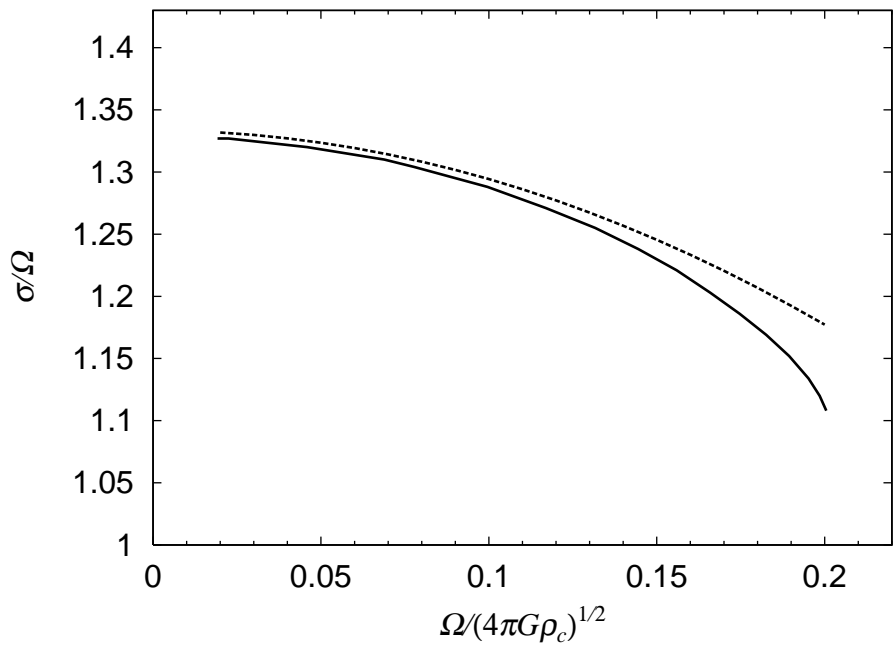
\end{document}